\documentclass{article}

\usepackage{arxiv}

\usepackage[utf8]{inputenc} 
\usepackage[T1]{fontenc}    
\usepackage{hyperref}       
\usepackage{url}            
\usepackage{booktabs}       
\usepackage{amsfonts}       
\usepackage{nicefrac}       
\usepackage{microtype}      
\usepackage{lipsum}
\usepackage{graphicx}
\graphicspath{ {./images/} }

\newcommand{\arcdeg}{$^{\circ}$}
\newcommand{\citet}{\cite}

\title{Gamma-Ray Burst Triangulation with a Near-Earth Network}

\author{
 Kevin C. Hurley \\
University of California \\
Space Sciences Laboratory \\
Berkeley, CA 94720-7450, USA \\
\texttt{khurley@ssl.berkeley.edu} \\
}

\begin{document}
\maketitle
\begin{abstract}
We study the characteristics of Near-Earth-Networks (NENs) of gamma-ray burst (GRB) detectors, with the objective of defining a network with all-sky, full-time localization capability for multi-messenger astrophysics.  We show that a minimum network consisting of 9 identical spacecraft in two orbits with different inclinations provides a good combination of sky coverage with several-degree localization accuracy with detector areas of 100 cm$^2$.  In order to achieve this, careful attention must be paid to systematics.  This includes accurate photon timing ($\sim$ 0.1 ms), good energy resolution ($\sim$ 10\%), and reduction of Earth albedo, which are all within current capabilities.  Such a network can be scaled in both the number and size of detectors to produce increased accuracy.  We introduce a new method of localization which does not rely on on-board trigger systems or on the cross-correlation of time histories, but rather, in ground processing, tests positions over the entire sky and assigns probabilities to them to detect and localize events.  We demonstrate its capabilities with simulations. If the NEN spacecraft can downlink at least several hundred time- and energy-tagged events per second, and the data can be ground-processed as they are received, it can in principle derive GRB positions in near-real time over the entire sky.
\end{abstract}


\section{Introduction} 
\label{sec:intro}

All-sky monitoring and localization of gamma-ray transients promises to be an important component of multi-messenger astrophysics.  Detections of gravitational waves, neutrinos, very high energy gamma-rays, and optical transients will occur at a combined rate of at least several per month for the foreseeable future, and inevitably the question of an associated gamma-ray burst (GRB) will arise for all of them.  In the case of LIGO/Virgo, gravitational wave-only localization areas are predicted to be up to 180 deg$^2$, with latencies of hours to days \citet{Abbott_2018}.  In other cases, precise localizations will become available almost in real time.  In all cases, however, it will be essential to have all-sky coverage in the $\sim$ 15 - 150 keV energy range.  Any single spacecraft in a low-Earth orbit cannot provide this capability, due to duty cycle, Earth-blocking, field-of-view considerations, or all three.  An average of 50\% coverage of points on the celestial sphere is a rough estimate of what can be achieved by a mission such as \it Fermi \rm, for example.  Although the current interplanetary network (IPN\footnote{http://ssl.berkeley.edu/ipn3},  \cite{doi:10.1063/1.3621810}) of gamma-ray burst detectors is functioning well and its spacecraft are not scheduled for de-commissioning, they are nevertheless old, and new interplanetary opportunities do not arise frequently.

Here we explore a different approach, based on networks of small, identical, dedicated GRB satellites in near-Earth orbits, with very simple instrumentation.  For example, these could be relatively simple 6U CubeSats, although that is not essential for this study.  We first discuss the capabilities of two possible Near Earth Networks (hereafter NENs), and show that a network consisting of nine spacecraft in two orbits with different inclinations provides the minimum acceptable coverage.  Then we introduce an innovative method of data analysis which can be used to derive localizations without time history comparisons, under the assumption of continuous time- and energy-tagged photon data transmission from multiple spacecraft with identical detection systems.  We demonstrate its capabilities with detailed simulations, and finally, discuss various means of enhancing the performance of the network.

We note that there are presently numerous alternative concepts such as GECAM, CAMELOT, and HERMES (among others), for all-sky monitoring using multiple Earth-orbiting satellites and a variety of localization techniques \cite{zheng_shijie_2019_3478126}, \cite{inproceedings1}, and \cite{inproceedings2}.

\section{A simple example} \label{sec:example}

\begin{figure*}
\includegraphics[scale=0.5]{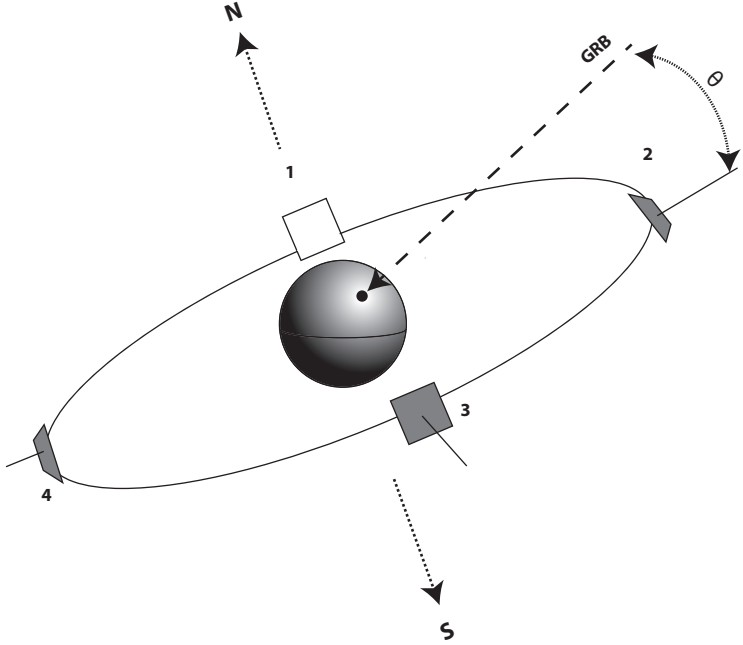}
\caption{A 4-spacecraft network.  The detectors are shielded below (white, detector 1) and their sensitive areas (gray, detectors 2, 3, and 4) have a planar geometry with axes pointed towards the spacecraft zeniths.  An incoming burst defines a sub-burst point at Earth (black dot).  Each detector effective area is determined by the GRB/zenith angle $ \theta$.  The north and south orbital poles are indicated.}
\label{fig:4scNEN}
\end{figure*}

We will consider a simple hypothetical NEN first.  When two spacecraft detect a GRB, an annulus of location is defined whose width is inversely proportional to the separation between the two spacecraft.  A third spacecraft gives a second annulus, which intersects the first one to define two roughly elliptical error regions.  To resolve the ambiguity, either a fourth, non-co-planar spacecraft is required, or geometrical considerations such as detector response as a function of arrival angle can be utilized.  We start with the assumption of N$_{sat}$ identical spacecraft (N$_{sat} \ge$ 3).  They are deployed along a circular orbit with identical altitudes A and inclinations i; the spacecraft have equal spacing (360\arcdeg /N$_{sat}$) along the orbit.  Thus they are co-planar.  Figure~\ref{fig:4scNEN} shows a 4-spacecraft network.  In these simulations, each spacecraft has a duty cycle d determined solely by South Atlantic Anomaly (SAA) passages.

For the simple network in this example, we assume a circular Earth with radius 6378 km.  We take  the altitude to be 600 km (roughly similar to that of \it Swift \rm or \it Fermi \rm), the inclination i to be 20$^{\circ}$, and the duty cycle d to be 0.85.  For reasons related to the effects of Earth albedo, explained in more detail in section~\ref{sec:differentapproach}, we assume that each spacecraft has a single, planar detector (i.e. with a cosine-law response), shielded on the sides and the Earth-facing plane, whose axis is pointed towards the local zenith.  In this simple example, each detector has unit area and can observe a burst up to 90$^{\circ}$ off-axis.

The quality and quantity of localizations in a network are a function of the number of spacecraft which can detect an event, and the total effective detection area.  Consider the number of spacecraft first, which we will refer to as the coverage.  For a single spacecraft at any given time, either a point on the sky is visible or it is not.  For a network, however, a point may be observable to a number of spacecraft, and the detection of an event by n detectors serves different purposes depending on n.  n=1 is a detection with no localization other than detector angular response considerations.  For n=2, triangulation to an annulus is possible, and angular response may limit the localization to portions of the annulus.  (The visible regions from both the detecting and the non-detecting spacecraft may be useful.)  For n=3, triangulation to two alternate error regions is possible, and angular response may eliminate one in some cases.  Since the spacecraft are co-planar in this example, the detection by more than 3 spacecraft may reduce the sizes of the alternate error boxes, but it does not remove their ambiguity, if triangulation is the only method employed. The total effective area is taken to be the sum of the individual detector effective areas.  

To calculate the coverage and total effective area, the sky was divided into 41273 equally spaced cells (each 1 square degree in area).  For the right ascension and declination of each cell, and for each spacecraft in the network, the cell/Earth center/spacecraft angle was calculated.  For angles less than 90$^{\circ}$, the cell was considered to be visible to the detector, and the detector area (the cosine of the angle) was calculated.  The duty cycle of each detector was assumed to be determined solely by SAA passages.  A spacecraft thus has a probability P of being in the SAA given by (1-d).  This in turn corresponds to a number of degrees along the orbit (1-d)*360$^{\circ}$. The calculation was repeated for N$_{sat}$=3, 4, 6, 8, and 10. For each network, a random point along the orbit was chosen for the first spacecraft, and the remaining spacecraft positions were calculated assuming uniform spacing along the orbit, keeping track of those in the SAA. Since each spacecraft observes half the sky, at most N$_{sat}$/2 detectors can observe any given right ascension and declination with non-zero area, before duty cycle considerations.  (Only the orbital poles can be observed by all the detectors, but with zero effective area.) 

\begin{figure*}
\includegraphics[scale=0.5]{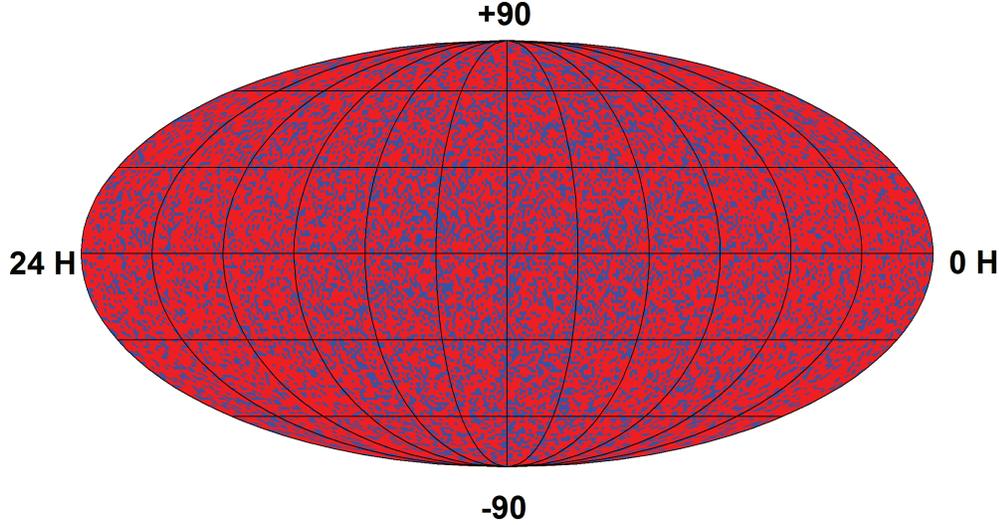}
\caption{The coverage (number of spacecraft which can observe a point on the sky) for a 4-spacecraft network.  Blue: 1 spacecraft.  Red: 2 spacecraft.  30\% of the sky is covered by one spacecraft, and 70\% is covered by 2 spacecraft.  Since there are no regions covered by 3 or 4 spacecraft, the localizations are at best annuli.}
\label{fig:4_600_20coverage}
\end{figure*}

\begin{figure*}
\includegraphics[scale=0.5]{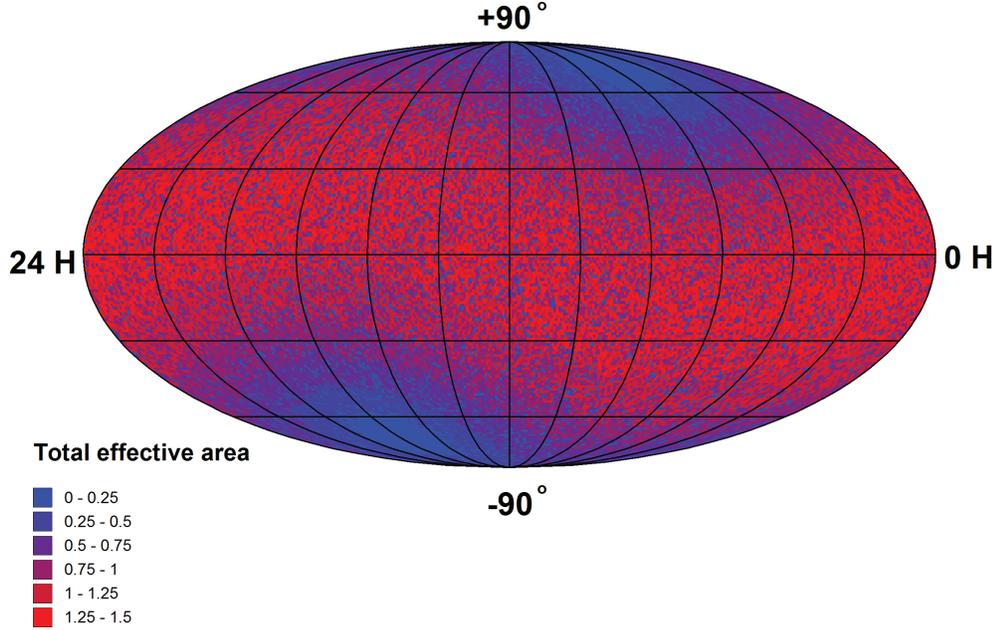}
\caption{The effective areas for points on the sky for a 4-spacecraft network.  Each spacecraft is assumed to have a detector with unit area.  The area goes to zero at the orbital poles.}
\label{fig:4_600_20area}
\end{figure*}

For a 3-spacecraft network, 51\% of the sky is observable to an average of 0.55 spacecraft, and 49\% is observable to an average of 1.55.  For networks of 4, 6, 8, and 10 spacecraft,  the entire sky (with the exception of the orbital poles) is observable to an average of 1.4, 2.1, 2.8, and 3.5 spacecraft, respectively. The coverage and total effective area as a function of right ascension and declination are shown in the sky maps of figure~\ref{fig:4_600_20coverage} and figure~\ref{fig:4_600_20area} for a network of 4 spacecraft.    

In all these networks, triangulation yields at best two alternate error boxes because the spacecraft are co-planar.  The error boxes lie at equal angles above and below the orbital plane.  Furthermore, because of the symmetry of the assumed detector pointing directions, each detector has identical responses to the two possible arrival directions, and the ambiguity cannot be resolved with detector response patterns. Another disadvantage of this configuration is that the sensitivity goes to zero at the orbital poles because all the detectors have zero effective area at these points.  These restrictions may be overcome by placing the spacecraft in orbits with different inclinations.

\section{A more complex example} 
\label{sec:morecomplex}

\begin{figure*}
\includegraphics[scale=1.0]{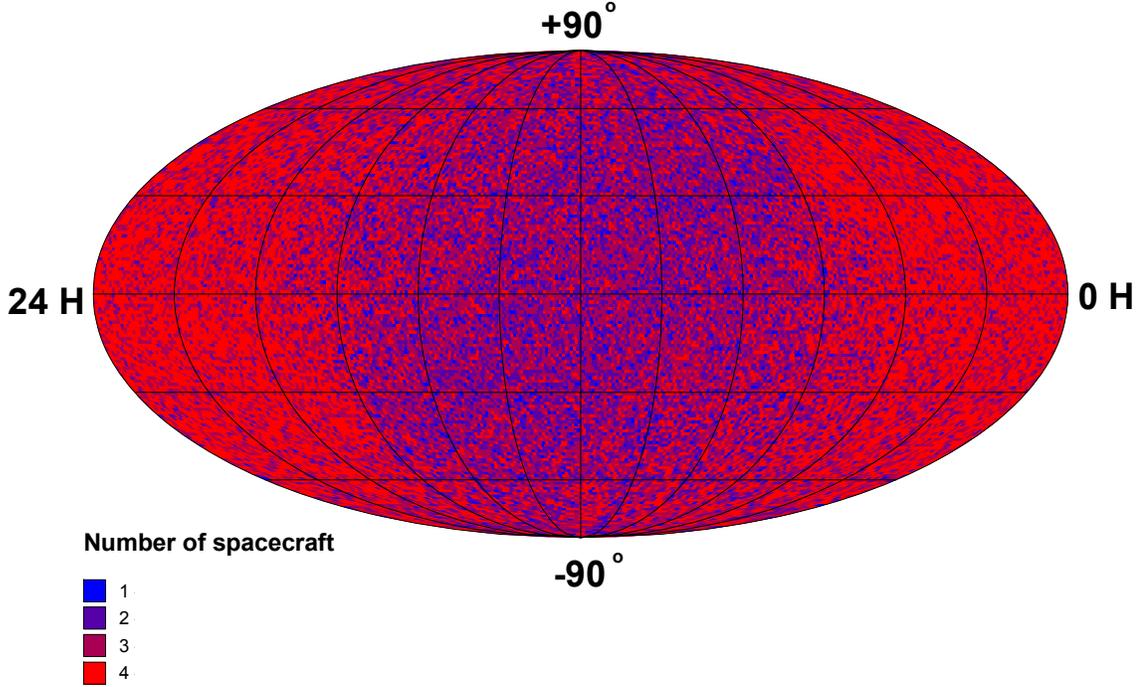}
\caption{The coverage (number of spacecraft which can observe a point on the sky) for a 9-spacecraft network.  3.5\%, 19.7\%, 38.8\%, 31.1\%, and 6.8\% of the sky are covered by one, two, three, four, and five spacecraft, respectively.  Thus localizations to annuli, double error boxes, and single error boxes are possible.}
\label{fig:9scnumberofsc}
\end{figure*}

\begin{figure*}
\includegraphics[scale=1.0]{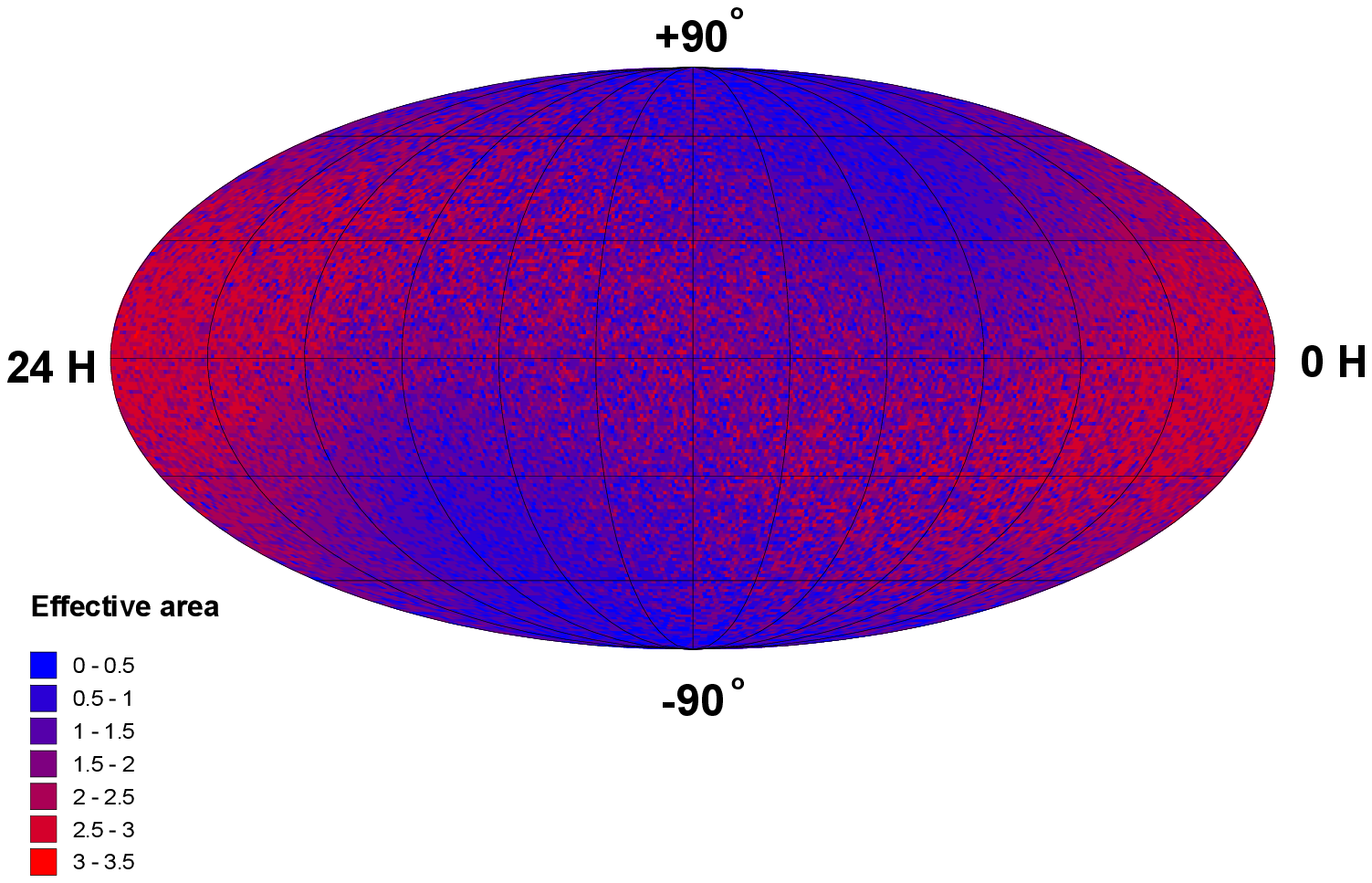}
\caption{The effective areas for points on the sky for a 9 spacecraft network.   Each spacecraft is assumed to have a detector with unit area.  The area decreases toward the two orbital poles, but because the spacecraft are in orbits with two different inclinations, it never goes to zero.}
\label{fig:9scarea}
\end{figure*}

Now consider a network consisting of spacecraft in two orbits with different inclinations, one equatorial (i=0$^{\circ}$) and one at i=51.6$^{\circ}$ (identical to the International Space Station).  The duty cycles in these orbits will be approximately 0.85 and 0.59.  This means that the coverage will be greater if more spacecraft are placed into the higher inclination orbit.  The effective area and coverage are shown in the sky maps of figure~\ref{fig:9scnumberofsc} and figure~\ref{fig:9scarea}.  Roughly 60\% of the sky is observed by 3, 2 or 1 spacecraft, leading to two error regions, an annulus, or no localization.  Roughly 40\% is observed by 4 or more spacecraft, leading to single error regions.  Because the orbital poles are separated by 51.6$^{\circ}$, the network has some sensitivity to every point on the sky. Because of the different inclinations, the symmetry of the preceding example is broken, which means that it is possible to obtain localizations to a single error box if the burst is observed by 4 or more spacecraft in 2 different orbits.  In the following sections, we describe a new procedure and detailed simulations to derive GRB error boxes with this network.

\section{A different approach to GRB localization} \label{sec:differentapproach}

For 40 years, GRB triangulation has been done by cross-correlating the time histories of spacecraft pairs to obtain annuli, and calculating their intersection points to define error boxes.  This has been driven by the fact that almost all detectors had trigger systems which provided high resolution, time-binned light curves, often in a single energy range, and only when a GRB was detected using a particular trigger algorithm (the trigger systems typically used only a few timescales to search for events).  With the exception of the earliest networks, most networks have contained an array of very different detectors, often with different time binning and different energy ranges, which introduced systematic uncertainties into light curve comparisons, necessitating careful calibrations of each spacecraft pair.  If we assume, however, that the NEN detectors are identical, and provide a continuous stream of time- and energy-tagged photons\footnote{it might be possible to forego energy-tagging if an active gain control system is used}, a better method, which does not involve time-binning or cross-correlation, can be used to search the entire sky continuously for transients on all time scales.

The first step of the method is to again divide the sky into a number N$_{cell}$ of equally spaced cells each centered at RA $\alpha$, declination $\delta$.  Then calculate the visibility of that $\alpha, \delta$ to each spacecraft in the network; there will be n visible spacecraft.  Calculate the time delays $\Delta$t for each visible spacecraft, with respect to the sub-$\alpha, \delta$ point, which can be taken as time zero.  For each visible spacecraft, calculate the number of counts on N$_{time}$ timescales in N$_e$ energy bins; the start time for each calculation is given by the delay, and the end time is determined by an assumed burst duration.  Subtract the background; since the count rates should follow a normal distribution, it is possible to assign a $\chi^2$ value to the sky bin, and a probability that a burst was detected.  Repeat the procedure for each sky bin to obtain a sky map in $\chi^2$ space, and search the map for significant detections.  

The threshold for a detection will depend on the total number of trials, N$_{cell}\times$N$_{time}\times$N$_{e}$, as well as the frequency of the searches.  N$_{cell}$ will be determined by the localization accuracy.  The useful timescales will be bounded at the lower end roughly by the durations of the shortest bursts observed to date.  The upper bound might be hundreds of seconds or longer.  N$_e$ will be determined by the detector energy range and energy resolution.  The maximum repetition rate, or frequency, can be set by the detector count rates (there is no point in repeating a sky search if no new photons have been recorded, for example).

This method avoids the disadvantages of time binning and cross-correlation, and it makes use of all the photons.  Computationally, however, it may be intensive.  For example, if we use 1$^{\circ}$ sky cells, N$_{cell}=4.1\times10^4$.  If we search for bursts with durations between 20 ms and 200 s on timescales that increase by a factor of two, N$_{time}$=13.  If we use 5 energy ranges, N$_e$=5, and a sky search involves $2.7\times10^6$ calculations.  If we want to be sure to detect short events (20 ms long), the search frequency might be around 50/second, which means $1.3\times10^8$ calculations/second.  There are several mitigating factors.  First, the calculations are relatively simple ones.  Second, the sky cell calculations can probably be done in parallel.  Finally, it is possible to use coarser sky cells and energy ranges initially, and finer ones if a burst is detected.

This method imposes two requirements on the detection system.  The first is accurate photon timing; 0.1 ms is adequate for these simulations, and achievable, and will be assumed; greater accuracies are possible and would reduce localization uncertainties further.  The second is an accurate comparison of the counts in the various detectors.  Atmospheric scattering (Earth albedo), in which burst photons reflected off the Earth’s atmosphere interact in the detectors, can be an important contribution to detector responses \cite{Pendleton1992}; indeed, \cite{Pendleton1999} showed that it could actually exceed the directly detected component in some cases for the BATSE detectors with the weakest direct response.  For BATSE and Fermi \cite{Connaughton2015}, atmospheric scattering corrections were done iteratively by first obtaining a coarse localization for the burst by comparing detector counts, but ignoring scattering.  Then using that localization, an atmospheric response database established for BATSE \cite{Pendleton1999} was consulted.  The database was constructed using a Monte Carlo code for 41,168 grid points, and for 3 types of burst energy spectra (soft, medium, and hard).  For each grid point and each spectral type, detector rates were calculated for the Earth-spacecraft geometry at the time of the trigger.  The atmospheric component was then added to the response.  A $\chi^2$ minimization was done for each of the three spectral types to find the most likely arrival direction in spacecraft coordinates.  The corrections were large enough that, as the missions progressed, improvements to the localization algorithm were largely the result of improved atmospheric scattering calculations.
This has two main implications for a NEN.  The first is that the scattered component will be delayed with respect to the direct component for each detector.  This adds uncertainties to the localization, which depends on summing counts over precise time intervals.  
The second is that, since relative detector responses are used to localize bursts, the detector responses would need to include scattered components.  In principle it might be possible to utilize a procedure similar to the BATSE/Fermi method  described above, with the added complication that the different detectors are on different spacecraft, and not at a single position.  However, in the following simulations, it has been assumed that the detector axes are all aligned towards their zeniths, and that the detectors are shielded along the Earth-facing, bottom plane and around the sides. It is further assumed that the detectors have good energy resolution to assure that count comparisons are in the same energy range for each detector.

\section{A detailed simulation.} \label{sec:simulation}

\begin{figure*}
\includegraphics[scale=0.5]{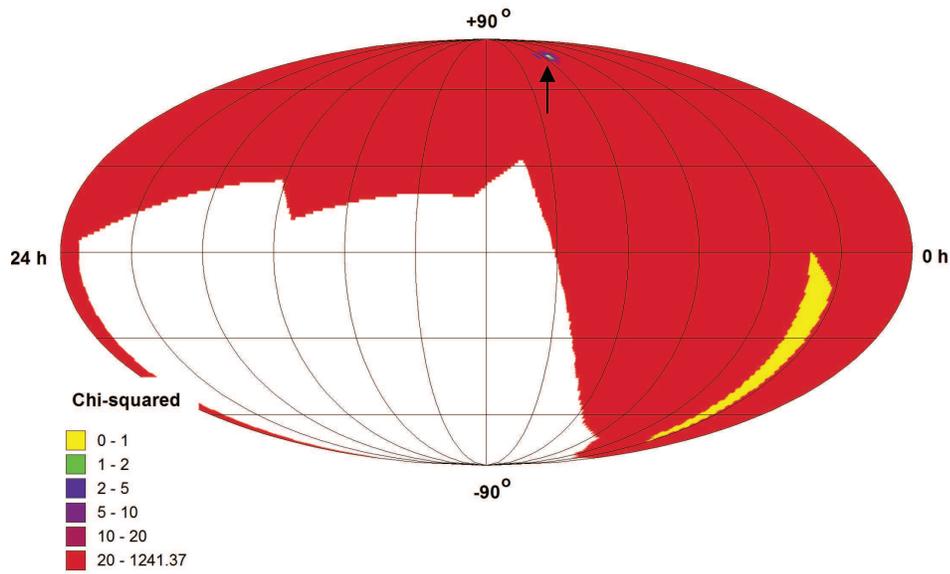}
\caption{Chi-squared map for a burst with 1400 counts at the sub-GRB point.  Only bursts detected by 4 or more spacecraft in 2 orbits were searched for.  The blank areas are points where fewer than 10 net counts were found in the detector nearest the sub-Earth point.  The arrow indicates the position of the simulated GRB.  When the number of degrees of freedom of each point is used to find the probabilities, the localization in figure~\ref{fig:contourfine} is obtained.}
\label{fig:nenchisquare}
\end{figure*}

\begin{figure*}
\includegraphics[scale=0.5]{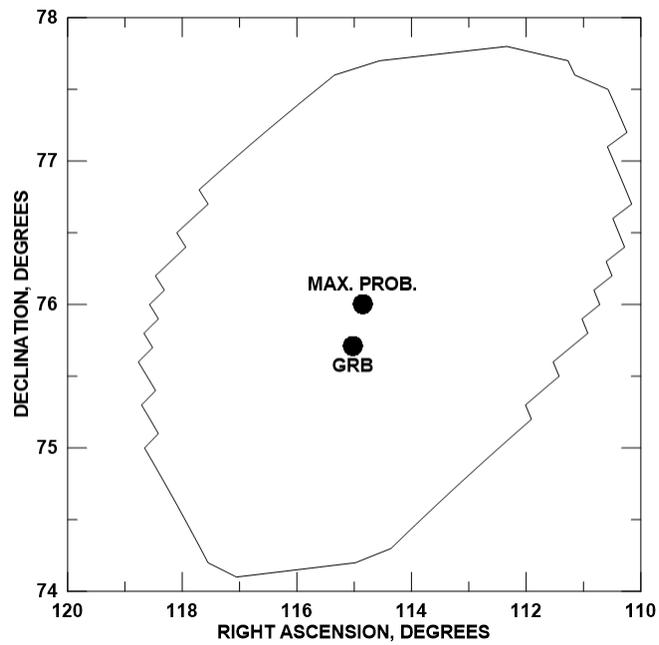}
\caption{The 3$\sigma$ localization for the $\chi^2$ map of figure~\ref{fig:nenchisquare}.  The $\chi^2$ values have been converted to probabilities using the number of degrees of freedom associated with each point on the map.  The simulated arrival direction of the GRB is indicated, as well as the point with the maximum probability.}
\label{fig:contourfine}
\end{figure*}

Simulations were carried out for the 9-spacecraft network described in Section \ref{sec:morecomplex}.  A random GRB arrival direction was selected, and the visibility of each detector in the network was calculated, taking field of view, duty cycle, and effective area into account.  For each detector, a time- and energy-tagged time history was constructed.  Photon arrival times were distributed according to Poisson interval statistics, with a steady background.  Time tags were taken to be accurate to 0.1 ms. The GRB time history was taken to be a simple top-hat (rectangular) function with a duration of 100 ms, with a variety of different intensities for different runs.  Since we want to compare the numbers of photons in different detectors with as much accuracy as possible, the detector energy resolution must be taken into account.  For any photon spectrum with an index less than zero, the energy resolution will cause more photons to leak into the measured energy band than out of it, at the lower energy threshold, and vice-versa at the higher threshold.  This results in different detectors counting different numbers of photons under otherwise identical conditions.  The photon spectrum was assumed here to be a simple power law with an index of -2.  The detector energy resolution (full-width at half-maximum) as a function of energy E was assumed to be 
$FWHM=A+B\times(E+C*E^2)^{0.5}$
where A=0.0059, B=0.0037, and C=8.9629, with E in MeV \cite{Ji2016}, typical of a CZT detector.  With the time histories assigned, a blind search for a GRB was carried out.  The first step involved constructing a single, co-added time history in a given energy range of all the detectors which were not off due to duty cycle considerations, and searching it for an increase in the count rates.  This identified the approximate start time and duration of the event, but not its arrival direction or the detectors which observed it.  In the next step, the sky was divided into 1$^{\circ}$ cells.  For each cell, detector arrival times were calculated relative to the sub-GRB point.  For each detector visible to the GRB and not off due to duty cycle, counts were integrated starting from its calculated GRB arrival time and lasting for the GRB duration.  If, after background subtraction, the detector nearest the sub-GRB point (and therefore with the greatest effective area) had more than 10 net counts, the number was corrected for the effective area to obtain a GRB intensity at the sub-GRB point.  The expected number of counts at the other detectors was calculated taking their effective areas into account and compared to the observed number.  Using all the detectors with more than 10 counts (except for the one nearest the sub-GRB point, which was used to determine the intensity), a $\chi^2$ was calculated, and a probability derived for the cell. In this procedure, the number of detectors, and therefore the number of degrees of freedom, may vary from cell to cell.  A $\chi^2$ sky map and its corresponding probability map are shown in figures~\ref{fig:nenchisquare} and ~\ref{fig:contourfine}. If an event was detected above a given probability threshold, the search area and the cell size were reduced, and the $\chi^2$ and probability calculations were repeated to obtain a more precise localization.

Depending on the assumed burst intensity, its arrival direction, and the number of detectors which observed it, the procedure outlined above produces probability maps which can define annuli, alternate error boxes, or single boxes. In the next step, we have chosen to concentrate on bursts detected by 4 or more spacecraft in two different orbits, which therefore produce single error boxes.

Three burst intensities were considered, corresponding to 140, 280, and 1400 net counts per detector on axis.  For a 100 cm$^2$ detector and an E$^{-2}$ spectrum, these correspond to fluences of approximately $10^{-7}\rm erg~cm^{-2}, 2\times10^{-7}\rm erg~cm^{-2}$, and $10^{-6}\rm erg~cm^{-2}$, and to peak fluxes of approximately $16~\rm photons~cm^{-2}~ s^{-1}$, $32~\rm photons~cm^{-2}~s^{-1}$,~and $160~\rm photons~cm^{-2}~s^{-1}$
in the 15 - 150 keV range.  1, 2, and 3$\sigma$ error boxes were derived for each burst intensity for 100 random arrival directions.  For each arrival direction, the simulation identified a point where the probability reached a maximum, and the number of points inside and outside the contours were counted to compare with their expected values from the confidence intervals.  Table~\ref{tab:areas} gives the average dimensions and areas of the localization regions for the above intensities and confidence intervals.

\begin{table}
\caption{Minimum and maximum dimensions, and areas of localization regions, at 1, 2, and 3$\sigma$ confidence levels, as a function of fluence in erg cm$^{-2}$ and counts per detector (in parentheses).  The numbers were derived from an average over 100 simulations in each case.}
\centering
\begin{tabular}{cccc}
\toprule
\label{tab:areas}
Fluence (counts) & 1 $\sigma$ & 2 $\sigma$ & 3 $\sigma$  \\
\midrule
$10^{-7}\rm erg~cm^{-2}$ (140)        &	4.5$^{\circ}$, 17.0$^{\circ}$, 64.7 sq. deg.  & 5.8$^{\circ}$, 24.7$^{\circ}$, 130.0 sq. deg. & 8.7$^{\circ}$, 37.3$^{\circ}$, 263.8 sq. deg. \\
$2\times10^{-7}\rm erg~cm^{-2}$ (280) &	3.5$^{\circ}$, 12.1$^{\circ}$, 36.0 sq. deg.  & 5.1$^{\circ}$, 17.2$^{\circ}$, 77.5 sq. deg.   & 6.9$^{\circ}$, 25.7$^{\circ}$, 141.5 sq. deg. \\
$10^{-6}\rm erg~cm^{-2}$ (1400)       & 1.6$^{\circ}$, 4.6$^{\circ}$,  6.7 sq. deg.   & 2.3$^{\circ}$, 7.8$^{\circ}$, 14.6 sq. deg.   & 3.2$^{\circ}$, 10.9$^{\circ}$, 28.9 sq. deg.  \\
\bottomrule
\end{tabular}
\end{table}
Figure~\ref{fig:nenchisquare} shows the $\chi^2$ map for one burst simulation, and figure~\ref{fig:contourfine} shows the resulting 3$\sigma$ contour.

\section{Discussion}

The NEN discussed in the previous sections achieves localization accuracies which are roughly comparable to those of the \it Fermi \rm GBM, but with the advantage of all-sky, full-time coverage.  Although this would be useful for multi-messenger astrophysics, it is not ideal, for several reasons. The first is that only about 40\% of bursts can be localized to a single error region.  The second is that the error regions are relatively large.   The first problem can be solved by adding more spacecraft.  If a total of 15 spacecraft are deployed into orbits with inclinations 0 and 51.6$^{\circ}$ (6 in the first orbit and 9 in the second), about 87\% of the sky can be covered by four or more detectors.  This also means that every point on the sky will be covered with greater effective area, but it does not substantially improve the localization accuracy, because as the number of spacecraft increases, their separation decreases, unless a 3rd orbit with a different inclination is used.  Greater localization accuracy can be achieved, however, with larger detector areas.  Table~\ref{tab:areas} demonstrates that the accuracy (error box dimensions or areas) is proportional to the reciprocal of the number of counts, so an increase in area of a factor of 10 would lead to sub-degree 1$\sigma$ dimensions for the brightest events.  Higher altitudes would increase the spacecraft separations and also reduce areas, but at the expense of higher backgrounds and reduced duty cycles.

\section{Conclusions}

Simulations demonstrate that gamma-ray burst localization with moderate accuracy can be achieved with a near-Earth network of simple detectors.  The advantages of such a network are simplicity, reduced cost, modularity, and full-time, all-sky coverage for events of practically any duration.  The challenges to operate a network such as this with the localization method in Section~\ref{sec:differentapproach} are careful control of systematics, high telemetry downlink rates, and substantial ground-processing capability.  None of these appears to be outside the scope of current or future capabilities.

Support from NASA Grant 80NSSC18K1722 is gratefully acknowledged.

\bibliographystyle{unsrt}  

\end{document}